\documentclass[12pt]{article}
\usepackage{hyperref}
\usepackage{amssymb}
\usepackage{graphicx}
\usepackage{amsmath}
\usepackage{authblk}
\usepackage[section]{placeins}
\numberwithin{equation}{section}
\title{Superdense star in a space-time with minimal length}
\author{Bhanu Kiran. S \thanks{bhanukiran.3696@gmail.com} , E. Harikumar \thanks{harisp@uohyd.ernet.in}, Vishnu Rajagopal \thanks{vishnurajagopal.anayath@gmail.com}}
\affil{\textit{School of Physics, University of Hyderabad,}\\ \textit{Central University P.O, Hyderabad-500046, Telangana, India}}
\date{}
\begin{document}
\maketitle
\begin{abstract}
In this paper we generalise core-envelope model of superdense star to a non-commutative space-time and study the modifications due to the existence of a minimal length, predicted by various approaches to quantum gravity. We first derive Einstein's field equation in $\kappa$-deformed space-time and use this to set up non-commutative version of core-envelope model describing superdense stars. We derive $\kappa$-deformed law of density variation, valid upto first order approximation in deformation parameter and obtain radial and tangential pressures in $\kappa$-deformed space-time. We also derive $\kappa$-deformed strong energy conditions and obtain a bound on the deformation parameter.
\end{abstract}

\section{Introduction}
A complete amalgamation of theory of general relativity and quantum mechanics is always a fascinating area in theoretical physics. Many recent 
studies in the area of quantum gravity predicts existence of a minimal length scale for the space-time framework. In this context 
non-commutative space-time finds to be an efficient way to handle the space-time structure at Planck scale \cite{connes,dop}. 

The influence of quantum gravity has been studied in many physical systems. Superdense star is one among them where the matter content in 
the star is tightly bound together, over a small region of space-time. When these regions become more and more dense, the effect of gravity 
increases and we expect the quantum gravity features to become significant. 

The Superdense stars are stars whose matter density distribution is much greater than the nuclear density. Neutron stars and white dwarfs are some of the typical examples for superdense stars. The study of dynamics of superdense star gives the relations between different thermodynamic variables. Such relations can be obtained in various ways such as by using statistical mechanics and quantum mechanics. The study of white dwarf as degenerate fermi electron gas \cite{3} is one such example. Similar study with the neutron stars predicted the existence of black holes \cite{4}. The most easy way to obtain the relations between different thermodynamic variables associated with stars is by deriving analyticaly the relevant equation of state. These equations of state can be obtained from Tolman-Oppenheineimer-Volkoff equation, which is derived from general theory of relativity. But when the matter density inside the star is much greater than the usual nuclear density, it is difficult to obtain a complete equation of state for the superdense matter distribution. Thus in this situation one uses an alternative approach called core-envelope models \cite{5,6,7}. In this approach a relativistic star is assumed to consists of a central core region surrounded by an outer envelope region, and the matter distribution in these regions exhibit different physical features. Plenty of works have been done using core-envelope models \cite{7, 8}. Most of these studies concluded that core and envelope regions have perfect fluid distributions, with different matter density, in equilibrium. Investigation reported in \cite{9} shows that both pressure and density are continous at the core-envelope boundary. Theoretical studies reported in \cite{10} and \cite{11} show that superdense matter distribution has an anisotropic pressure distribution, which degenerates into radial and tangential pressures. Some models with constant matter density and spherical anisotropic distributions were studied in \cite{12}. Later in \cite{13}, the anisotropic distribution model with variable matter density was studied. Core-envelope models with anisotropic core as well as isotropic envelope and vice-versa were studied in \cite{14} and \cite{15}, respectively.

In this paper we study the behaviour of superdense stars, assuming their geometry is non-commutative. We introduce the non-commutativity through non-commutative metric and non-commutative generalisation of energy momentum tensor, thereby setting up non-commutative version of Einstein equation. Here we will use the core-envelope model of superdense stars, with core having isotropic and envelope having anisotropic pressure distributions. Further, we consider the core and envelope regions to possess a perfect fluid distribution under equlibrium conditions. 

The non-commutativity in space-time introduces a fundamental length scale, which is known to arise in the context of quantum gravity. Moyal space-time is one such space-time exhibiting a non-commutative geometry. This space-time coordinates satisfy the commutation relation $[\hat{x}_{\mu}, \hat{x}_{\nu}]=\Theta_{\mu\nu}$, where $\Theta_{\mu \nu}$ is a constant tensor. A detailed behaviour of black holes in Moyal space-time has been studied in \cite{16}. In the present work, we will use another non-commutative space-time, called $\kappa$-deformed space-time, whose coordinates obey a Lie algebra type commutation relation between the space and time coordinates \cite{17, 18, 19, 20,1,2}. Theoretical studies of black hole in $\kappa$-space-time have been reported in \cite{21, 22, 23} and properties of compact stars in $\kappa$-deformed space-time is studied \cite{24}.

Different physical models on $\kappa$-space-time have been studied using various approaches. One way is to express the non-commutative functions in terms of commutative coordinates and then study thus obtained commutative theory. In this work we will use this method where we expresses the non-commutative coordinates as a function of commutative coordinates, their derivatives and deformation parameter. Different realisations of non-commutative coordinates exist for $\kappa$-deformed space-time \cite{25, 26, 27, 28, 29}. Here we have choosen a particular realisation for calculations.

Organisation of this paper is as follows. In section 2 we give a brief summary of the superdense star \cite{30}. One start with the 4 dimensional flat space metric to which a 3-spheroid space is embedded. This mertic is parameterised by defining two geometric parameters, which play a crucial role in developing the core-envelope model for superdense star, and obtain the static spherically symmetric metric, representing superdense matter. We then construct the energy momentum tensor for superdense matter distribution by considering it as a perfect fluid with an additional anisotropic term. Using the metric and energy momentum tensor, we will write down the Einstein field equation. In section 3, we choose a specific realisation for $\kappa$-deformed phase space. This helps us in writing the non-commutative variables in terms of commutative variables, its derivatives and deformation parameter. Using this, we obtain $\kappa$-deformed metric that represent superdense matter. Next we construct the $\kappa$-deformed energy momentum tensor, and using this deformed metric and deformed energy momentum tensor, we formulate the $\kappa$-deformed Einstein field equation. For all these cases the commutative limits are recovered by setting the $\kappa$-deformation parameter to zero. From the deformed Einstein equation, we derive the $\kappa$-deformed law of density variation. By demanding the positivity of deformed density, we get a bound on the $\kappa$-deformation parameter. Similarly, we get a bound on deformation parameter by demanding the deformed density gradient to be negative. In section 4, we solve the deformed Einstein field equation explicitly in the isotropic core and the anistropic envelope, separately. Their solutions enable us to write down the $\kappa$-deformed central pressure in the core, $\kappa$-deformed radial and tangential pressures in the envelope, and the $\kappa$-deformed strong energy conditions. From the positivity condition for the deformed tangential pressure, we set another bound on $\kappa$-deformation parameter. Apart from this, we also get the expression for mass of the star using the metric continuity condition at the boundary of superdense star. By introducing the compactness factor, using mass, we calculate the surface redshift. These informations are used to plot the variation of quantities such as deformed density, deformed pressure (both radial and tangential), derivative of deformed pressure with respect to deformed density against the star radius. Finally in section 5, we present our results and conclusions drawn from them.
\section{Static spherically symmetric metric}
In this section, we briefly summerise the derivation of the metric appropriate for representing superdense stars. We also discuss the construction of energy momentum tensor for the same. This will enable one to write down the Einstein field equations for superdense stars, which is used in the derivation of equation of states.

A 3-spheroid embedded in the 4-dimensional Euclidean flat space of metric
\begin{equation}
d\sigma^2=dx^2+dy^2+dz^2+dw^2, 
\end{equation}
is described by the equation.
\begin{equation} \frac{w^2}{b^2}+\frac{x^2+y^2+z^2}{R^2}=1. \end{equation}
Using the parametrisation of the 3-spheroid \cite{30}
\begin{align}
x&=R\sin\psi\sin\theta\cos\phi, \\
y&=R\sin\psi\sin\theta\sin\phi, \\
z&=R\sin\psi\cos\theta, \\
w&=b\cos\psi,
\end{align}
the Eq.(2.1) is re-written as
\begin{equation}
d\sigma^2=(R^2\cos^2\psi+b^2\sin^2\psi)d\psi^2+R^2\sin^2\psi(d\theta^2+\sin^2\theta d\phi^2). \end{equation}
Defining $r=R\sin\psi$, and introducing a geometric parameter $K=1+\frac{b^2}{R^2} >1$, one rewrites the 4-dimensional Euclidean metric as
\begin{equation}d\sigma^2=\frac{\Big(1+K\frac{r^2}{R^2}\Big)}{\Big(1+\frac{r^2}{R^2}\Big)}dr^2+r^2\Big( d\theta^2+\sin^2\theta d\phi^2 \Big). \end{equation}
Here $K$ and $R$ represent the geometric parameters.
Using this we write the space-time metric as
\begin{equation} ds^2=e^{f(r)}dt^2-d\sigma^2. \end{equation}
The above metric takes the familiar static, spherically symmetric form given by
\begin{equation} ds^2=e^{f(r)}dt^2-e^{\lambda (r)}dr^2-r^2(d\theta^2+\sin^2\theta d\phi^2) \end{equation}
with
\begin{equation}e^{\lambda(r)}=\frac{1+K\frac{r^2}{R^2}}{1+\frac{r^2}{R^2}}. \end{equation}
From the line element given in Eq.(2.10), we read off the metric components, $g_{\mu\nu}$. This metric has been used for studying superdense matter distributions \cite{31, 32}. 
We use perfect fluid distribution with an anistropic term to model the superdense matter distribution. Thus the energy momentum tensor is given by
\begin{equation} T_{\mu\nu}=(\rho+p)u_{\mu}u_{\nu}-pg_{\mu\nu}+\Pi_{\mu\nu}. \end{equation}
This energy momentum tensor has a radially symmetric anisotrpic fluid distribution, if the dynamical quantities have the form \cite{33}.
\begin{align}
\rho&=\rho(r) \\
p&=p(r)\\
\Pi _{\mu\nu}&=\sqrt{3}S(r)\bigg[C_{\mu} C_{\nu}-\frac{1}{3}(u_{\mu}u_{\nu}-g_{\mu\nu})\bigg]
\end{align}
Here $S(r)$ denotes the magnitude of the anisotropic stress tensor and $u_{\mu}$ denotes the unit four velocity of the fluid matter. One takes without loss of generality, $u_{\mu}=(e^{f(r)/2},0,0,0)$  which is a choice for the equilibrium condition and $C_{\mu}$ denotes the radial vector. We also choose $C_{\mu}=(0,e^{\lambda(r) /2},0,0)$.

 Now radial and tangential pressure are given by
\begin{equation} p_r=p+\frac{2S(r)}{\sqrt 3}, \end{equation}
\begin{equation} p_t=p-\frac{S(r)}{\sqrt 3}, \end{equation}
respectively. In terms of $p_r$ and $p_t$ , the anistropic stress tensor is given to be
\begin{equation} S(r)=\frac{p_r-p_t}{\sqrt 3}. \end{equation}
The curvature of the space-time (determined from metric) is connected to matter content in space-time by Einstein field equations. Here we write the Einstein field equations for the case where cosmological constant is zero, i.e $\Lambda=0$, and thus we have
\begin{equation} G_{\mu\nu}=8\pi T_{\mu\nu}, \end{equation}
where $G_{\mu\nu}=R_{\mu\nu}-\frac{1}{2}Rg_{\mu\nu}$ is the Einstein tensor. $R_{\mu\nu}$ and $R$ are Ricci tensor and Ricci scalar, respectively. Thus by calculating $R_{\mu\nu}$ and $R$ for the metric given in Eq.(2.10) and using Eq.(2.12), one writes down the Einstein equations explicitly. The solution to these equations give the relation between various quantities of interest.
\section{$\kappa$-deformed Einstein field equations}
In this section, first we summerise the construction of metric in the $\kappa$-deformed space-time \cite{2}. Then we use deformed metric as well as deformed energy-momentum tensor and derive $\kappa$-deformed Einstein field equations, which form the basis of our analysis. Following this, we study the variation of density with respect to distance from origin. These results will be used in the next section in analysing superdense star in non-commutative space-time. 

The coordinates of $\kappa$-deformed space-time obey the commutation relations
\begin{equation} [\hat{x}_0,\hat{x}_i]=ia\hat{x}_i, ~~ [\hat{x}_i, \hat{x}_j]=0, \end{equation}
where $a$ is the deformation parameter having the dimension of length.

We assume that coordinates of $\kappa$-deformed phase space obey the commutation relation \cite{25, 34, 35},
\begin{equation} [\hat{x}_{\mu},\hat{P}_{\nu}]=i\hat{g}_{\mu\nu}, \end{equation} 
where $\hat{g}_{\mu\nu}$ is the $\kappa$-deformed metric. Now we introduce another $\kappa$-deformed space-time coordinate $\hat{y}_{\mu}$ such that it commutes with $\hat{x}_{\mu}$ i.e $[\hat{y}_{\mu},\hat{x}_{\nu}]=0$. The new coordinate is introduced only for calculational simplification. 

We choose a specifc realization of $\kappa$-deformed phasespace \cite{25, 34, 35} given by
\begin{equation} \hat{x}_{\mu}=x_{\alpha}\varphi^{\alpha}_{\mu}, \,\hat{P}_{\mu}=g_{\alpha\beta}(\hat{y})k^{\beta}\varphi^{\alpha}_{\mu}. \end{equation}
From Eq.(3.1) and Eq.(3.2) we get
\begin{equation} \varphi _0^0=1, \, \varphi _i^0=0, \, \varphi_0^i=0, \, \varphi _j^i=\delta _j^i e^{-ak^0}. \end{equation}
The coordinate $\hat{y}$ also satisfies the commutation relation given in Eq.(3.1). We demand that $\hat{y}_{\mu}$ also can be written in terms of commutative coordinate and momenta as
\begin{equation} \hat{y}_{\mu}=x_{\alpha}\phi_{\mu}^{\alpha}(p). \end{equation}
Using Eq.(3.5) and $[\hat{x}_{\mu},\hat{y}_{\nu}]=0$ we find
\begin{equation} \hat{y}_0=x_0-ax_jk^j, \end{equation}
\begin{equation} \hat{y}_i=x_i. \end{equation}
The $\kappa$-deformed metric can be obtained \cite{2} from Eq.(3.2) as
\begin{equation} [\hat{x}_{\mu},\hat{P}_{\nu}] \equiv i\hat{g}_{\mu\nu}=ig_{\alpha\beta}(\hat{y})\Big(k^{\beta}\frac{\partial \varphi^{\alpha}_{\nu}}{\partial k^{\sigma}}\varphi_{\mu}^{\sigma}+\varphi_{\mu}^{\alpha}\varphi_{\nu}^{\beta}\Big). \end{equation}
Here $g_{\mu\nu}(\hat{y})$ is obtained by replacing the commutative coordinates with the $\kappa$-deformed coordinates in the commutative metric and therefore we have $g_{\alpha\beta}(\hat{y})=g_{\beta\alpha}(\hat{y})$.

Using Eq.(3.2) and Eq.(3.4) in Eq.(3.8), we find \cite{2}
\begin{align} [\hat{x}_0,\hat{P}_0]&=ig_{00}(\hat{y}), \\
 [\hat{x}_0,\hat{P}_i]&=ig_{i0}(\hat{y})\big(1-ak^0\big)e^{-ak^0}-ag_{ik}(\hat{y})k^ke^{-ak^0},\\
 [\hat{x}_i,\hat{P}_0]&=ig_{i0}(\hat{y})e^{-ak^0},\\
 [\hat{x}_i,\hat{P}_j]&=ig_{ij}(\hat{y})e^{-2ak^0}
\end{align}
Using Eq.(3.9)-(3.12) we get the explicit form of $\hat{g}_{\mu\nu}$ as
\begin{align}
\hat{g}_{00}&=g_{00}(\hat{y}),\\
\hat{g}_{0i}&=g_{i0}(\hat{y})\big(1-ak^0\big)e^{-ak^0}-ag_{im}(\hat{y})k^me^{-ak^0}, \\
\hat{g}_{i0}&=g_{0i}(\hat{y})e^{-ak^0}, \\
\hat{g}_{ij}&=g_{ij}(\hat{y})e^{-2ak^0}.
\end{align}
The line element in $\kappa$-deformed space-time is defined  as \cite{2}
\begin{equation}d\hat{s}^2=\hat{g}_{\mu\nu}d\hat{x}^{\mu}d\hat{x}^{\nu}. \end{equation}
Thus the explicit form of the line element is given by
\begin{multline}d\hat{s}^2=g_{00}(\hat{y})dx^0dx^0+\Big(g_{i0}(\hat{y})\big(1-ak^0\big)-ag_{im}(\hat{y})k^m\Big)e^{-2ak^0}dx^0dx^i+\\  g_{0i}(\hat{y})e^{-2ak^0}dx^idx^0+g_{ij}(\hat{y})e^{-4ak^0}dx^idx^j. \end{multline}
Now we use the above procedure and generalise the metric in Eq.(2.10) to $\kappa$-space-time. From Eq.(3.7) we see that $g_{\mu\nu}(\hat{y}_i)=g_{\mu\nu}(x_i)$. Since the cross terms in time and space indices of metric in Eq.(2.10) are zero, the $\kappa$-deformed metric given in Eq.(3.18) reduces in our case to
\begin{equation}d\hat{s}^2=g_{00}(\hat{y})dx^0dx^0+g_{ij}(\hat{y})e^{-4ak^0}dx^idx^j. \end{equation}
From Eq.(2.10), we read off $g_{00}(\hat{y})$ and $g_{ij}(\hat{y})$ and using this, the resulting $\kappa$-deformed space-time metric corresponding to one given in Eq.(2.10) is obtained as
\begin{equation} \label{b} d\hat{s}^2=e^{f(r)}dt^2-\big[e^{\lambda (r)}dr^2+r^2(d\theta^2+\sin^2\theta d\phi^2)\big]e^{-4ak^0}. \end{equation}
Thus the above equation defines the $\kappa$-deformed metric corresponding to static, spherically symmetric space-times.
Here we see that only the spatial parts of the metric get modified under $\kappa$-deformation by an overall factor of $e^{-4ak^0}$ . In the limit $a \to 0$, we recover the metric in commutative case, given in Eq.(2.10).

In the $\kappa$-deformed space-time, the energy momentum tensor is taken to be
\begin{equation}\label{a} \hat{T}_{\mu\nu}=(\rho+p)\hat{u}_{\mu}\hat{u}_{\nu}-p\hat{g}_{\mu\nu}+\hat{\Pi}_{\mu\nu}, \end{equation}
where $\hat{u}_{\mu}=u_{\alpha}\varphi_{\mu}^{\alpha}$ and $\hat{C}_{\mu}=C_{\alpha}\varphi_{\mu}^{\alpha}$. From Eq.(3.4) we get $\hat{u}_{\mu}=(e^{f(r)/2},0,0,0)$ and $\hat{C}_{\mu}=(0,e^{\lambda(r)/2}e^{-ak^0},0,0)$.
We find that only the diagonal components of Eq.(3.21) survives and thus we have
\begin{align}
\hat{T}_{00}&=\rho e^{f(r)}, \\
\hat{T}_{11}&=\Big(pe^{-4ak^0}+\frac{2S}{\sqrt 3}e^{-ak^0} \Big)e^{\lambda(r)},\\
\hat{T}_{22}&=\Big(p-\frac{S}{\sqrt 3}\Big)r^2e^{-4ak^0}, \\
\hat{T}_{33}&=\Big(p-\frac{S}{\sqrt 3}\Big)r^2\sin ^2\theta e^{-4ak^0}.
\end{align}
Here we see that only the spatial components i.e, $\hat{T}_{ij}$  get modified under $\kappa$-deformation whereas the temporal component remains unchanged i.e, $\hat{T}_{00}=T_{00}$.

Under the $\kappa$-deformation, the Einstein tensor becomes 
\begin{equation} \hat{G}_{\mu\nu}=\hat{R}_{\mu\nu}-\frac{1}{2}\hat{R}\hat{g}_{\mu\nu} \end{equation} 
where $\hat{R}_{\mu\nu}$ and $\hat{R}$ are $\kappa$-deformed Ricci tensor and Ricci scalar respectively. The non vanishing components of $\kappa$-deformed Ricci tensor calculated using the $\kappa$-deformed metric defined by Eq.(3.20) are
\begin{multline*}\hat{R}_{00}=\frac{\Big(R^2f'^2(r)+2R^2rf''(r)+4R^2f'(r)+Kr^3f'^2(r)\Big)R^2e^{f(r)}e^{4ak^0}}{4r(R^4+2R^2Kr^2+K^2r^4)}+\\
\frac{\Big(R^2rf'^2(r)+2R^2rf''(r)+6R^2f'(r)+Kr^3f'^2(r)\Big)r^2e^{f(r)}e^{4ak^0}}{4r(R^4+2R^2Kr^2+K^2r^4)}+\\\frac{\Big(2R^2Kr^3f''(r)+2Kr^5f''(r)+2R^2Kr^2f'(r)+4Kr^4f'(r)\Big)e^{f(r)}e^{4ak^0}}{4r(R^4+2R^2Kr^2+K^2r^4)},\end{multline*}
\begin{multline*}\hat{R}_{11}=\frac{2R^2r(R^2+r^2)(K-1)f'(r)-(R^2+Kr^2)(R^4+2R^2r^2+r^4)f'^2(r)}{4(R^2+Kr^2)(R^4+2R^2r^2+r^4)}+\\
\frac{8R^2(R^2+r^2)(K-1)-2(R^2+Kr^2)(R^4+2R^2r^2+r^4)f''(r)}{4(R^2+Kr^2)(R^4+2R^2r^2+r^4)},\end{multline*}
\begin{multline*}\hat{R}_{22}=\frac{-rR^2\Big(R^2f(r)+Kr^2f(r)-4Kr+r^2f'(r)\Big)}{2R^4+4R^2Kr^2+2K^2r^4}\\+\frac{-r^2\Big(4R^2-2K^2r^2+Kr^3f'(r)+2Kr^2\Big)}{2R^4+4R^2Kr^2+2K^2r^4},\end{multline*}
\begin{equation*}\hat{R}_{33}=\hat{R}_{22}\sin^2\theta.\end{equation*}
Now we determine the components of Einstein field tensor from Eq.(3.26) as
\begin{equation}\hat{G}_{00}=\frac{\bigg(3KR^2-3R^2+K^2r^2-Kr^2\bigg)e^{f(r)}e^{4ak^0}}{R^4+2KR^2r^2+K^2r^4},\end{equation}
\begin{equation}\hat{G}_{11}=\frac{f'(r)R^2-Kr+r^2f'(r)+r}{r(R^2+r^2)}, \end{equation}
\begin{multline}\hat{G}_{22}=\frac{r\bigg(R^4rf'^2(r)+2r^4f''(r)+2R^4f'(r)+R^2Kr^3f^2(r)\bigg)}{4(R^4+2R^2Kr^2+K^2r^4)}+\\\frac{r\bigg(R^2r^3f'^2(r)+2R^2r^3f''(r)+4R^2r^2f'(r)+4R^2r)\bigg)}{4(R^4+2R^2Kr^2+K^2r^4)}+\\\frac{r\bigg(2R^2Kr^3f''9r)-4R^2Kr+Kr^5f'^2(r)+2Kr^4f'(r)\bigg)}{4(R^4+2R^2Kr^2+K^2r^4)},\end{multline}
\begin{equation}\hat{G}_{33}=\sin^2\theta\hat{G}_{22}.\end{equation}
Note that only the $\hat{G}_{00}$ component gets modified by the $\kappa$-deformation whereas the spatial components remain unchanged, unlike $\hat{T}_{\mu\nu}$.

In the $\kappa$-space-time, the Einstein field equation Eq.(2.20) takes the form
\begin{equation} \hat{G}_{\mu\nu}=8\pi\hat{T}_{\mu\nu}. \end{equation}
From this we obtain three independent $\kappa$-deformed Einstein field equations, valid upto the first order approximation in the deformation parameter $a$, as
\begin{equation}8\pi\rho\Big(1-4ak^0\Big)=\frac{(K-1)\Big(3+K\frac{r^2}{R^2}\Big)}{R^2\Big(1+K\frac{r^2}{R^2}\Big)}, \end{equation}
\begin{equation}8\pi\Big(p_r-4ak^0-\frac{2S}{\sqrt 3}ak^0\Big)=\frac{\Big(1+\frac{r^2}{R^2}\Big)\frac{f'(r)}{r}-\frac{(K-1)}{R^2}}{1+K\frac{r^2}{R^2}}, \end{equation}
\begin{multline}8\pi S\sqrt 3\Big(1-2ak^0\Big)=\frac{r(K-1)}{R^2}\bigg(\frac{1}{r}+\frac{f'(r)}{2}\bigg)\bigg(1+K\frac{r^2}{R^2}\bigg)^{-2}-\frac{(K-1)}{R^2}\bigg(1+K\frac{r^2}{R^2}\bigg)^{-1}\\-\bigg(\frac{f''(r)}{2}+\frac{f'^2(r)}{4}-\frac{f'(r)}{2r}\bigg)\bigg(1+\frac{r^2}{R^2}\bigg)\bigg(1+K\frac{r^2}{R^2}\bigg)^{-1}.
\end{multline}
We note that all these field equations get modified in $\kappa$-space-time. This is due to the contribution of $\hat{G}_{00}$ and $\hat{T}_{ij}$ components, respectively.
\subsection{$\kappa$-deformed law of density variation}
In this subsection we derive the law of density variation in $\kappa$-deformed superdense star using the above results.

Note the $a$ dependent correction term obtained above in Eq.(3.32). We obtain the non-commutative density by bringing the $(1-4ak^0)$ term to the right hand side and applying binomial expansion, keeping upto the first order approximation in $a$ (here we denote non-commutative density with a hat as $\hat{\rho}$). Thus $\kappa$-deformed density is given as
\begin{equation}\hat{\rho}=\frac{(K-1)\Big(3+K\frac{r^2}{R^2}\Big)}{8\pi R^2\Big(1+K\frac{r^2}{R^2}\Big)^2}\Big(1+4ak^0\Big). \end{equation}
Here we see that the density gets modified by a factor $1+4ak^0$. We recover the well known result in commutative limit when $a\to 0$. 
For $\hat{\rho}$ to be postive, we need to choose $K>1$ and $ak^0>-0.25$, respectively. In SI units this becomes $\frac{ak^0}{ch}>-0.25$. If we choose $k^0$ as Planck energy (i.e $k^0\approx 10^{19}$ GeV), we obtain a lower bound on the deformation parameter $|a|\geq 10^{-36}$m.

From the above equation we find the $\kappa$-deformed central density (denoted by $\hat{\rho}_0=\hat{\rho}(0)$, i.e, $\hat{\rho}$ at $r=0$) as
\begin{equation}\hat{\rho_0}=\frac{3(K-1)}{8\pi R^2}\Big(1+4ak^0\Big).\end{equation}
On the boundary of star where $r=r_2$, density attains the value
\begin{equation} \hat{\rho}(r_2)=\frac{(K-1)\Big(3+K\frac{r_2^2}{R^2}\Big)}{8\pi R^2\Big(1+K\frac{r_2^2}{R^2}\Big)^2}\Big(1+4ak^0\Big). \end{equation}
Thus we observe that as $r$ increases, $\hat{\rho}$ decreases from a maximum value $\hat{\rho}_0$ at the centre to $\hat{\rho}(r_2)$ on the boundary and we have shown this graphically in Fig.\ref{fig:density}. This behaviour is exactly the same as the behaviour of $\rho$ in commutative case.

Note that $\kappa$-deformed density gradient 
\begin{equation} \frac{d\hat{\rho}}{dr}=\frac{-2K(K-1)r}{8\pi R^4}\frac{\big(5+K\frac{r^2}{R^2}\big)}{\big(1+K\frac{r^2}{R^2}\big)^3}\Big(1+4ak^0\Big) \end{equation}
is negative for $K>1$ and $|a|\geq 10^{-36}$m.

For a particular choice $K=2$, Eq.(3.35), Eq.(3.36) and Eq.(3.38) become
\begin{equation}\hat{\rho}=\frac{(3+2\frac{r^2}{R^2}\big)}{8\pi R^2\big(1+2\frac{r^2}{R^2}\big)}\Big(1+4ak^0\Big), \end{equation}
\begin{equation}\hat{\rho_0}=\frac{3}{8\pi R^2}\Big(1+4ak^0\Big), \end{equation}
\begin{equation} \frac{d\hat{\rho}}{dr}=\frac{-r}{2\pi R^4}\frac{\big(5+2\frac{r^2}{R^2}\big)}{\big(1+2\frac{r^2}{R^2}\big)^3}\Big(1+4ak^0\Big). \end{equation}
In the remaining calculations we choose the geometric parameter $K$ to be $2$.
\section{Superdense star}
In this section we analyse the superdense star in the space-time with a minimal length (introduced through non-commutativity). We use the results of previous sections in setting up the equations of state relevant for this scenario. We use the core-envelope model, generalised to $\kappa$-deformed space-time to study superdense star in the non-commutative setting. Thus we first set up $\kappa$-deformed Einstein field equations for the isotropic core. This is followed by same analysis for anisotropic envelope.
\subsection{Isotropic Core}
In this subsection, we obtain the $\kappa$-deformed Einstein field equations for isotropic fluid distribution in the core.

The core of the star is in the region $0\leq r \leq r_{1}$. Here the anisotropic stress tensor $S(r)$ vanishes hence the radial pressure and tangential pressure become equal.\\
Inside the core, Eq.(3.34) becomes (with $K=2$)
\begin{multline}\frac{r}{R^2}\bigg(\frac{1}{r}+\frac{f'(r)}{2}\bigg)-\frac{1}{R^2}\bigg(1+2\frac{r^2}{R^2}\bigg)-\bigg(\frac{f''(r)}{2}+\frac{f'^2(r)}{4}-\frac{f'(r)}{2r}\bigg)\bigg(1+\frac{r^2}{R^2}\bigg)\bigg(1+2\frac{r^2}{R^2}\bigg)=0.
\end{multline}
Now by choosing $z=\sqrt{1+\frac{r^2}{R^2}}$ and $F=e^{f(r)/2}$, Eq.(4.1) becomes
\begin{equation} (2z^2-1)\frac{d^2F}{dz^2}+2z\frac{dF}{dz}-2F=0, \end{equation}
whose solution is given by \cite{15},
\begin{equation} F=A\sqrt{1+\frac{r^2}{R^2}}+B\bigg(\sqrt{1+\frac{r^2}{R^2}}L(r)-\frac{1}{\sqrt 2}\sqrt{1+2\frac{r^2}{R^2}}\bigg).\end{equation}
Here A, B are constants and 
\begin{equation} L(r)=\textrm{ln}\bigg(\sqrt 2\sqrt{1+\frac{r^2}{R^2}}+\sqrt{1+2\frac{r^2}{R^2}}\bigg). \end{equation}
Using these in Eq.(3.20), we obtain the $\kappa$-deformed metric inside the core to be
\begin{multline} d\hat{s}^2=F^2(r)dt^2-e^{\lambda (r)}dr^2e^{-4ak^0}-r^2(d\theta^2+\sin^2\theta d\phi^2)e^{-4ak^0}. \end{multline}
From Eq.(3.33) we thus find the $\kappa$-deformed core pressure to be 
\begin{equation} \hat{p}=\frac{\Bigg[A\sqrt{1+\frac{r^2}{R^2}}+B\bigg[\sqrt{1+\frac{r^2}{R^2}}L(r)+\frac{1}{\sqrt 2}\sqrt{1+2\frac{r^2}{R^2}}\bigg]\Bigg]}
{8\pi R^2(1+2\frac{r^2}{R^2})\bigg[A\sqrt{1+\frac{r^2}{R^2}}+B\Big[\sqrt{1+\frac{r^2}{R^2}}L(r)-\frac{1}{\sqrt 2}\sqrt{1+2\frac{r^2}{R^2}}\Big]\bigg]}\Big(1+4ak^0\Big). \end{equation} 
Here we see that the $\kappa$-deformed pressure in the core gets scaled by $1+4ak^0$ factor as compared to the core pressure in commutative case. When $a\to 0$, we obtain the core pressure in commutative limit. 

At the centre of the star the $\kappa$-deformed pressure is
\begin{equation} \hat{p}_0=\frac{\Big(A+B\big(L(0)+\frac{1}{\sqrt 2}\big)\Big)}
{8\pi R^2\Big(A+B\big(L(0)-\frac{1}{\sqrt 2}\big)\Big)}\Big(1+4ak^0\Big). \end{equation} 
From Eq.(3.35), Eq.(3.36), Eq.(4.6) and Eq.(4.7) we see
\begin{equation}\frac{\hat{\rho}}{\hat p}=\frac{\rho}{p}.\end{equation}
Thus we see that the ratio of density and pressure (in core) of the star in $\kappa$-deformed space-time is same as that in the commutative space. 

The $\kappa$-deformed strong energy condition 
\begin{equation} \hat{\rho}-3\hat{p}>0 \end{equation}
is satisfied inside the core if $A, B$ and $L(r)$ (see Eq.(4.3))satisfy the condition
\begin{equation} \frac{A+BL(r)}{B}>\frac{\Big(4+3\frac{r^2}{R^2}\Big)\sqrt{1+2\frac{r^2}{R^2}}}{2\sqrt 2\sqrt{1+\frac{r^2}{R^2}}}. \end{equation}
 Thus the $\kappa$-deformed density and pressure also obey the deformed strong energy condition, as that in the commutative limit\cite{31}.

 Using Eq.(3.41) and Eq.(4.6) the derivative of $\kappa$-deformed pressure with respect to $\kappa$-deformed density is obtained to be
\begin{equation}\frac{d\hat{p}}{d\hat{\rho}}=\frac{\Big(1+2\frac{r^2}{R^2}\Big)}{\Big(5+2\frac{r^2}{R^2}\Big)}+\frac{\sqrt 2 BR^2F'(r)\Big(1+2\frac{r^2}{R^2}\Big)^{5/2}}{4rF^2(r)\Big(5+2\frac{r^2}{R^2}\Big)}+\frac{\sqrt 2B\big(1+2\frac{r^2}{R^2}\big)^{3/2}}{F(r)\Big(5+2\frac{r^2}{R^2}\Big)}. \end{equation}
In Fig.\ref{fig:variation}, we plot derivative of $\hat{p}$ with respect to $\hat{\rho}$ against $\frac{r^2}{R^2}$ and have shown numericaly that it is less than one, i.e $\frac{d\hat{p}}{d\hat{\rho}}<1$. Thus we see that velocity of propagation of sound in isotropic core is less than velocity of light and it is in agreement with the commutative result of \cite{31}.
\subsection{Anisotropic Envelope}
In this subsection the envelope of the star defined as the region $r_{1}\leq r \leq r_{2}$, is analysed. The envelope is taken to be anisotropic (i.e, $S\neq 0$) and hence its radial pressure and tangential pressure are different from each other.

As in \cite{15}, we choose 
\begin{equation}\psi=\frac{e^{f(r)/2}}{(2z-1)^{1/4}}, z=\sqrt{1+\frac{r^2}{R^2}},\end{equation}and re-express Eq.(3.34) as
\begin{equation}\frac{d^2\psi}{dz^2}+\bigg(\frac{3(2z^2-1)-5z^2}{2z^2-1}+\frac{8\sqrt 3\pi R^2 S(1-2ak^0)(2z^2-1)}{z^2-1}\bigg)\psi=0. \end{equation} 
In order to simplify the calculation, we demand the terms in the bracket to be zero. This gives the $\kappa$-deformed anisotropic stress tensor (valid upto first order approximation in deformation parameter $a$) as
\begin{equation}\hat{S}=\frac{\frac{r^2}{R^2}\Big(2-\frac{r^2}{R^2}\Big)\Big(1+2ak^0\Big)}{8\pi \sqrt 3 R^2\Big(1+\frac{r^2}{R^2}\Big)^3}. \end{equation}
Here we recover anisotropic stress tensor in commutative case in lim $a \to 0$. Unlike the $\kappa$-deformed density and core pressure, the anisotropic stress tensor gets modified by a 
factor $1+2ak^0$. Hence Eq.(4.13) becomes
\begin{equation} \frac{d^2\psi}{dz^2}=0 \end{equation}
whose solution is given by
\begin{equation} \psi=Cz+D. \end{equation}
Here C and D are constants. From Eq.(4.12) and Eq.(4.15), we get
\begin{equation} e^{f(r)/2}=\Big(1+2\frac{r^2}{R^2}\Big)^{1/4}\Big(C\sqrt{1+\frac{r^2}{R^2}}+D\Big). \end{equation}
Thus in the envelope of the superdense star, $\kappa$-deformed metric becomes 
\begin{equation} d\hat{s}^2=\sqrt{1+2\frac{r^2}{R^2}}\Big(C\sqrt{1+\frac{r^2}{R^2}}+D\Big)^2dt^2-\frac{1+2\frac{r^2}{R^2}}{1+\frac{r^2}{R^2}}dr^2e^{-4ak^0}-r^2d\Omega ^2e^{-4ak^0}. \end{equation}
Using Eq.(3.33) and Eq.(4.14) the $\kappa$-deformed radial pressure is found to be
\begin{equation} \hat{p}_r=\frac{\Bigg[C\sqrt{1+\frac{r^2}{R^2}}\Big(3+4\frac{r^2}{R^2}\Big)+D\Bigg]\Big(1+4ak^0\Big)}{8\pi R^2\Big(1+2\frac{r^2}{R^2}\Big)^2\Big(C\sqrt{1+\frac{r^2}{R^2}}+D\Big)}-\frac{2\frac{r^2}{R^2}\Big(2-\frac{r^2}{R^2}\Big)ak^0}{8\pi R^2\Big(1+2\frac{r^2}{R^2}\Big)^3}. \end{equation}
Thus we have an extra term (second term) in radial pressure that appears due to $\kappa$-deformation apart from the modification due to $1+4ak^0$ factor (for density and core pressure). We obtain the commutative radial pressure when $a\to 0$.

 The deformed radial pressure, $\hat{p}_r$ is positive throughout the envelope if we choose $r\geq\sqrt 2R$ and $ak^0\geq 0$.
 
 Similarly, using Eq.(2.18), Eq.(4.14) and Eq.(4.19) the $\kappa$-deformed tangential pressure is found to be
\begin{equation}\hat{p}_t=\frac{\Bigg[C\sqrt{1+\frac{r^2}{R^2}}\Big(3+4\frac{r^2}{R^2}\Big)+D\Bigg]\Big(1+4ak^0\Big)}{8\pi R^2\Big(1+2\frac{r^2}{R^2}\Big)^2\Big(C\sqrt{1+\frac{r^2}{R^2}}+D\Big)}-\frac{\frac{r^2}{R^2}\Big(2-\frac{r^2}{R^2}\Big)\Big(1+4ak^0\Big)}{8\pi R^2\Big(1+2\frac{r^2}{R^2}\Big)^3}. \end{equation}
Note that tangential pressure also gets modified by a factor $1+4ak^0$. In the limit $a \to 0$, we recover commutative tangential pressure.
 
 Here also we see that $\hat{p}_t$ is positive throughout the envelope if we choose $r\geq\sqrt 2R$ and $1+4ak^0 \geq 0$.
 
Now we consider a typical superdense neutron star whose central density, envelope radius and core radius are $\rho_0=11.1145X10^{17} \textrm{kg/m}^3$, $r_1=11.330X10^3$m and $r_2=12.527X10^3$m, respectively \cite{36}. Using the relation $r_2^2> 2R^2$ and the expression for the $\kappa$-deformed central density i.e, $8\pi\rho(1-4ak^0)=\frac{3}{R^2}$ we set a bound on the deformation parameter given to be  $|a|>10^{-16}$m. 

The deformed strong energy condition in the envelope 
\begin{equation} \hat{{\rho}}-\hat{p}_r-2\hat{p}_t>0 \end{equation}
is satisfied if the constants C and D satisfy
\begin{equation}
\frac{D}{C}>\frac{\bigg(3\Big(1-3\frac{r^2}{R^2}+8\frac{r^4}{R^4}\Big)+2ak^0\Big(6+14\frac{r^2}{R^2}+29\frac{r^4}{R^4}\Big)\bigg)}{\bigg(\Big(1+4\frac{r^2}{R^2}+2\frac{r^4}{R^4}\Big)+2ak^0\Big(2+3\frac{r^2}{R^2}+12\frac{r^4}{R^4}\Big)\bigg)}\sqrt{1+\frac{r^2}{R^2}}. \end{equation}
This is similar to the strong energy condition as that in the commutative limit. 

From Eq.(4.19) and Eq.(4.20) we see that deformed tangential pressure is greater than the deformed radial pressure throughout the envelope. Using this in Eq.(2.18), we find that anisotropic factor is a negative quantity throughout the envelope. So from Eq.(4.14), we get the relation $(2-\frac{r^2}{R^2})(1+2ak^0)\leq 0$. From above arguments we find $r^2\geq 2R^2$, thus we obtain a bound on deformation parameter given by $|a|\geq 10^{-36}$m.

Using Eq.(3.41), Eq.(4.19) and Eq.(4.20) the derivative of $\kappa$-deformed radial and tangential pressures with respect to $\kappa$-deformed density are given by
\begin{multline}\frac{d\hat{p}_r}{d\hat{\rho}}=\frac{2\Big(1+2\frac{r^2}{R^2}\Big)+2\Big(1-5\frac{r^2}{R^2}+\frac{r^4}{R^4}\Big)ak^0}{\Big(1+2\frac{r^2}{R^2}\Big)\Big(5+2\frac{r^2}{R^2}\Big)}-\frac{C\Big(1+2\frac{r^2}{R^2}\Big)^2}{2\sqrt{1+\frac{r^2}{R^2}}\Big(5+2\frac{r^2}{R^2}\big)\psi(r)}+\\
\frac{CR^2\sqrt{1+\frac{r^2}{R^2}}\big(1+2\frac{r^2}{R^2}\Big)\bigg[\Big(1+2\frac{r^2}{R^2}\Big)\psi'(r)+4\frac{r^2}{R^2}\psi(r)\bigg]}{2r\Big(5+2\frac{r^2}{R^2}\Big)\psi'(r)}, \end{multline} and
\begin{multline}\frac{d\hat{p}_t}{d\hat{\rho}}=\frac{2\frac{r^2}{R^2}\Big(1-\frac{r^2}{R^2}\Big)}{\Big(5+2\frac{r^2}{R^2}\Big)\Big(1+2\frac{r^2}{R^2}\Big)}-\frac{C\Big(1+2\frac{r^2}{R^2}\Big)^2}{2\sqrt{1+\frac{r^2}{R^2}}\Big(5+2\frac{r^2}{R^2}\big)\psi(r)}+\\
\frac{CR^2\sqrt{1+\frac{r^2}{R^2}}\big(1+2\frac{r^2}{R^2}\Big)\bigg[\Big(1+2\frac{r^2}{R^2}\Big)\psi'(r)+4\frac{r^2}{R^2}\psi(r)\bigg]}{2r\Big(5+2\frac{r^2}{R^2}\Big)\psi'(r)}. \end{multline} 
In Fig.\ref{fig:variation} we plot this against $\frac{r^2}{R^2}$ and have shown numerically that both the quantities are less than one, i.e $\frac{d\hat{p}_r}{d\hat{\rho}}<1$ and $\frac{d\hat{p}_t}{d\hat{\rho}}<1$. Thus we see that velocity of propagation of sound in anisotropic envelope is less than velocity of light and it is in agreement with \cite{31}. The deformed superdense distribution is stable since $\frac{d\hat{p}_r}{d\hat{\rho}}>\frac{d\hat{p}_t}{d\hat{\rho}}$ \cite{37}.

Now we determine the constants $C$ and $D$ by matching the $\kappa$-deformed metric in the envelope with $\kappa$-deformed Schwarzchild exterior metric given by
\begin{equation}d\hat{s}^2=\Big(1-\frac{2m}{r}\Big)dt^2-\frac{1}{\Big(1-\frac{2m}{r}\Big)}e^{-4ak^0}dr^2-r^2e^{-4ak^0}d{\Omega}^2. \end{equation}
Equating Eq.(4.17) and Eq.(4.25) at $r=r_2$ we get
\begin{equation} m=\frac{r_2^3}{2R^2\Big(1+2\frac{r_2^2}{R^2}\Big)}, \end{equation}
\begin{equation} C\sqrt{1+\frac{r_2^2}{R^2}}+D=\frac{\sqrt{1+\frac{r_2^2}{R^2}}}{\Big(1+2\frac{r_2^2}{R^2}\Big)^{3/4}}. \end{equation}

At the boundary $r=r_2$, the $\kappa$-deformed radial pressure vanishes. i.e, $\hat{p}_r(r_2)=0$. This will give
\begin{equation} 8\pi\hat{p}_r(r_2)=\frac{\Bigg[C\sqrt{1+\frac{r_2^2}{R^2}}\Big(3+4\frac{r_2^2}{R^2}\Big)+D\Bigg]\Big(1+4ak^0\Big)}{R^2\Big(1+2\frac{r_2^2}{R^2}\Big)^2\Big(C\sqrt{1+\frac{r_2^2}{R^2}}+D\Big)}-\frac{2\frac{r_2^2}{R^2}\Big(2-\frac{r_2^2}{R^2}\Big)ak^0}{R^2\Big(1+2\frac{r_2^2}{R^2}\Big)^3}=0. \end{equation}
By solving Eq.(4.27) and Eq.(4.28) we obtain $C$ and $D$ as
\begin{equation} C=\frac{-\bigg(\Big(1+2\frac{r_2^2}{R^2}\Big)\Big(1+4ak^0\Big)-2\frac{r_2^2}{R^2}\Big(2-\frac{r_2^2}{R^2}\Big)ak^0\bigg)}{2\Big(1+2\frac{r_2^2}{R^2}\Big)^2\Big(1+4ak^0\Big)\Big(1+2\frac{r_2^2}{R^2}\Big)^{3/4}}, \end{equation}
\begin{equation} D=\sqrt{1+\frac{r_2^2}{R^2}}\frac{\bigg(\Big(3+4\frac{r_2^2}{R^2}\Big)\Big(1+2\frac{r_2^2}{R^2}\Big)\Big(1+4ak^0\Big)-2\frac{r_2^2}{R^2}\Big(2-\frac{r_2^2}{R^2}\Big)ak^0\bigg)}{2\Big(1+2\frac{r_2^2}{R^2}\Big)^2\Big(1+4ak^0\Big)\Big(1+2\frac{r_2^2}{R^2}\Big)^{3/4}}. \end{equation}
Thus we find that the $\kappa$-deformation of density and pressure modifies the constants $C$ and $D$ also. Even though $C$ and $D$ get modified, the quantity $\Big(C\sqrt{1+\frac{r_2^2}{R^2}}+D\Big)$ remains unchanged so that the $\hat{g}_{00}$ component of the metric on the boundary of superdense star remains unchanged under the $\kappa$-deformation.

At the core-envelope boundary, the anisotropy vanishes and hence $\hat{p}(r_1)_{core}=\hat{p}_r(r_1)=\hat{p}_t(r_1)$. The constants $A$ and $B$ can be determined from the continuity of pressure and metric coefficients at $r=r_1$. This gives the following two linear equations in $A$ and $B$ as
\begin{align}
\sqrt 3 A+B[\sqrt 3 L(r_1)+\sqrt{2.5}]&=\frac{11\sqrt 3 C +D}{5^{3/4}}, \\
\sqrt 3 A+B[\sqrt 3 L(r_1)-\sqrt{2.5}]&=\frac{\sqrt 3 C +D}{5^{3/4}}.
\end{align}
By solving Eq.(4.31) and Eq.(4.32) we get
\begin{align}
A&=\frac{[5\sqrt 5-3\sqrt 2(\sqrt 3L(r_2)-\sqrt{2.5})]C+\frac{1}{\sqrt 3}[5 \sqrt 5+2\sqrt 2(\sqrt 3L(r_1)-\sqrt{2.5})]D}{5^{3/4}}, \\
B&=\frac{\sqrt 2[3\sqrt 3C-2D]}{5^{3/4}}.
\end{align}
Now we define the compactness factor which gives the mass-to-radius ratio. It is defined as \cite{38, 39, 40}
\begin{equation}u=\frac{m(r_2)}{r_2}. \end{equation}
From Eq.(4.26) the compactness factor is found to be 
\begin{equation}u=\frac{m(r_2)}{r_2}=\frac{r_2^2}{2R^2\Big(1+2\frac{r_2^2}{R^2}\Big)}. \end{equation}
Using compactness factor the surface redshift is defined as (see \cite{38, 40} for details)
\begin{equation}\mathcal{Z}_R=\frac{1}{\sqrt{1-2u}}-1. \end{equation}
By substituting Eq.(4.36) in Eq.(4.37) the surface redshift for superdense star is
\begin{equation}\mathcal{Z}_R=\sqrt{\frac{1+2\frac{r_2^2}{R^2}}{1+\frac{r_2^2}{R^2}}}-1. \end{equation}
Here we notice that the surface redshift does not get modified under $\kappa$-deformation. This is because of that fact that the $\hat{g}_{00}$ component of deformed metric remains unchanged on the surface of superdense star under $\kappa$-deformation. 
\section{Discussions and Conclusions}
In this paper we have derived the $\kappa$-deformed Einstein field equation for a superdense star. We solved the field equations separately for core and envelope of superdense star upto first order approximation in $a$.
The Eq.(3.35) and Eq.(3.38) give the $\kappa$-deformed law of density variation, which is exaclty similar to the law of density variation in commutative space. From the conditions $\hat{\rho}>0$ and $\frac{d\hat{\rho}}{dr}<0$, we have obtained a bound on the deformation parameter given by $|a|>10^{-36}$m. From Eq.(4.6) we obtain the deformed isotropic core pressure, which is scaled by a factor $(1+4ak^0)$ as compared to the commutative case. Similarly from Eq.(4.19) and Eq.(4.20), we read off the deformed radial and tangential pressures, respectively. Here also we notice that the deformed tangential pressure gets scaled by $(1+4ak^0)$ factor. The positivity condition for tangential pressure and the expression for central density yield a limit on deformation parameter given by $|a|>10^{-16}$m and this is in agreement with the bound obtained in \cite{2}. The negative value of bound on deformation parameter $a$ is seen in \cite{41} also.  

We have obtained the deformed strong energy conditions in the core and envelope for $\kappa$-deformed superdense star, which is in a similar form as that in the commutative case \cite{30, 31}.

We have derived the surface compactness factor and surface redshift for superdense star in $\kappa$-deformed space-time, upto first order approximation in $a$, which are given by Eq.(4.36) and Eq.(4.38), respectively. The surface redshift remains unchanged under $\kappa$-deformation. This is because the $\hat{g}_{00}$ component of the $\kappa$-deformed metric on the boundary of $\kappa$-deformed superdense star do not get any $a$ dependent correction.

\begin{figure}[tbh]
\includegraphics[width=0.8\textwidth]{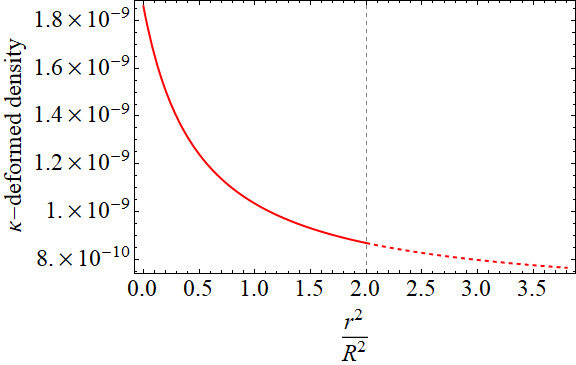}
\caption{Variation of $\hat{\rho}$ against $\frac{r^2}{R^2}$ throughout the star $(0\leq\frac{r^2}{R^2}\leq 3.8)$}
\label{fig:density}
\end{figure}

\begin{figure}[tbh]
\includegraphics[width=1.1\textwidth]{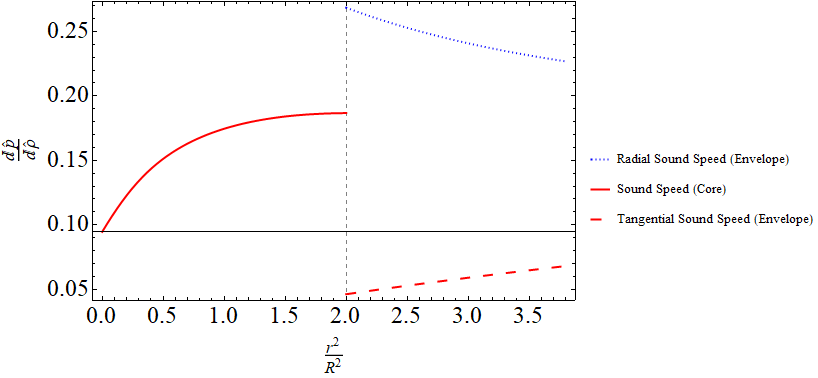}
\caption{Variation of $\frac{d\hat{p}}{d\hat{\rho}}$ against $\frac{r^2}{R^2}$ in the core $(0 \leq \frac{r^2}{R^2} \leq 2)$ and variation of $\frac{d\hat{p}_r}{d\hat{\rho}}$ and $\frac{d\hat{p}_t}{d\hat{\rho}}$ against $\frac{r^2}{R^2}$ in the envelope $(2 \leq \frac{r^2}{R^2} \leq 3.8)$.}
\label{fig:variation}
\end{figure}

\begin{figure}[thb]
\includegraphics[width=1.1\textwidth]{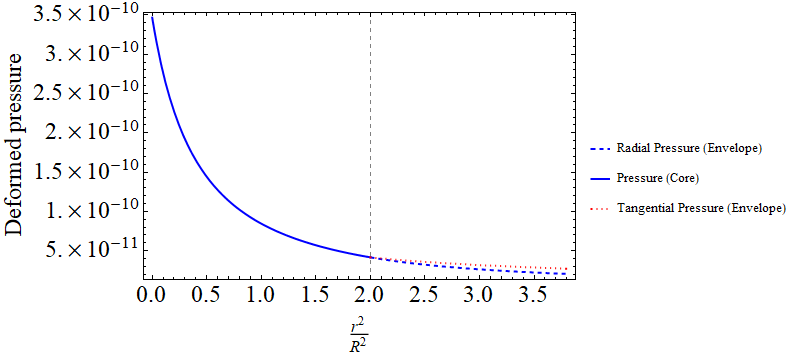}
\caption{Variation of $\hat{p}$ against $\frac{r^2}{R^2}$ in the core $(0 \leq \frac{r^2}{R^2} \leq 2)$ and variation of $\hat{p}_r$ and $\hat{p}_t$ against $\frac{r^2}{R^2}$ in the envelope $(2 \leq \frac{r^2}{R^2} \leq 3.8)$.}
\label{fig:pressure}
\end{figure}

 In Fig.\ref{fig:density}, Fig.\ref{fig:variation} and Fig.\ref{fig:pressure} we plot the variation of $\kappa$-deformed density, derivative of $\kappa$-deformed pressure with respect to $\kappa$-deformed density and $\kappa$-deformed pressure against $\frac{r^2}{R^2}$ (in natural units), respectively. Here we choose $ak^0 \sim 0.0001$. In Fig.\ref{fig:density} we see that deformed density decreases from a maximum value at centre $(\hat{\rho}_0)$ to a minimum value at boundary of the star, which is exactly similar to the behaviour of density in commutative case \cite{15}. This similarity in behaviour of density is due to the fact that the non-commutative correction is encoded only through scaling by $(1+4ak^0)$ factor. From Fig.\ref{fig:variation}, we see that the relations $\frac{d\hat{p}}{d\hat{\rho}}<1$ (inside the core), $\frac{d\hat{p}_r}{d\hat{\rho}}<1$ and $\frac{d\hat{p}_t}{d\hat{\rho}}<1$ (inside the envelope) are satisfied throughout the core-envelope, which implies that the speed of sound does not exceed the speed of light in the distribution and this is in agreement with commutative case \cite{31}. We also find that deformed radial speed of sound is greater than the deformed tangential speed of sound. From Fig.\ref{fig:pressure}, we observe that the deformed tangential pressure is greater than the deformed radial pressure throughout the envelope. This is also in agreement with the behaviour of radial and tangential pressure in commutative case \cite{15}, as expected.  
\subsection*{\bf Acknowledgments}
One of us (EH) thanks SERB, Govt. of India, for support through EMR/2015/000622. VR thanks Govt. of India, for support through EMR/2015/000622
and DST-INSPIRE/IF170622. BKS thanks P. C. Vinodkumar for useful discussions and encouragement.
 \FloatBarrier

\end{document}